\documentclass{aastex631}
\usepackage{upgreek}
\begin{document}

\title{On the fate of slow boulders ejected after DART impact on Dimorphos }

\author[0000-0003-0670-356X]{Fernando Moreno}
\affiliation{Instituto de Astrof\'isica de Andaluc\'ia, CSIC \\
Glorieta de la Astronom\'ia, s/n, 18008 Granada, Spain}
\author[0000-0002-4943-8623]{Gonzalo Tancredi}
\affiliation{Departamento de Astronom\'ia, Facultad de Ciencias, Igu\'a 4225, 11400 Montevideo, Uruguay}
\author[0000-0001-9840-2216]{Adriano Campo Bagatin}
\affiliation{
Instituto de F\'isica Aplicada a las Ciencias y las Tecnolog\'ias (IUFACyT),  Universidad de Alicante, \\ San Vicent del Raspeig, 03690 Alicante, Spain}
\affiliation{Departamento de F\'isica, Ingenier\'ia de Sistemas y Teor\'ia de la Se\~nal \\ Universidad de Alicante, San Vicent del Raspeig, 03690 Alicante, Spain}




\begin{abstract}

On 2022 September 26th, 23:14 UT the NASA/DART (Double Asteroid Redirection Test) spacecraft successfully impacted Dimorphos, the secondary component of the binary (65803) Didymos system, demonstrating asteroid orbit deflection for the first time. A large amount of debris, consisting on a wide size frequency distribution of particulates (from micron-sized dust to meter-sized boulders), was released, and a long-lasting tail has been observed over more than 9 months since impact. An important fraction of the ejecta mass has been ejected as individual meter-sized boulders, as have been found in images obtained by the Light Italian CubeSat for Imaging of Asteroid (LICIACube), as well as from the Hubble Space Telescope (HST). While the boulders observed by LICIACube had projected speeds of several tens of meter per second, those seen by the HST were about one hundred time slower. In this paper we analyze the long-term orbital evolution of those slow boulders using different dynamical codes, providing constraints on the fate of such large particles, and giving insight on the possibility of observing some of those boulders that might remain in orbit at the time of the ESA/Hera mission arrival to the binary system in late 2026.

\end{abstract}

\keywords{minor planets, asteroids: individual (65803 Didymos) --- methods: numerical}

\section{Introduction}\label{sec:intro}

The Double Asteroid Redirection Test (DART) is a NASA spacecraft that successfully impacted Dimorphos, the secondary member of the system (65803) Didymos \citep{2023Natur.616..457C, 2023Natur.616..443D}. The impact occurred on 2022 September 26th, 23:14:24.183 UT. DART impacted in a nearly head-on configuration on Dimorphos surface \citep{2023Natur.616..443D}, giving rise immediately to a fast ejected material (plume) (speed of $\approx$2 km s$^{-1}$) showing a spectrum characterized by emission lines of ionized alkali metals (NaI, KI, and LiI) \citep{2023Icar..40115595S}. This plume was clearly observed on images obtained from Les Makes Observatory  \citep{2023Natur.616..461G}, and was also seen in the earliest images during the HST monitoring \citep{2023Natur.616..452L}. A wide  ejection cone of dust particles and meter-sized boulders was monitored by the Light Italian CubeSat for Imaging of Asteroid \citep[LICIACube,][]{2021P&SS..19905185D,
2023LPICo2806.2426F} which separated from the DART spacecraft before the impact, performed a fast flyby of the system. Those boulders were found to move at speeds of several 10 m/s \citep{2023LPICo2806.2426F}. Ground-based and HST observations of the ejecta revealed complex structures with an initial diffuse dust cloud that evolved rapidly into a cone-shaped morphology and a solar radiation pressure driven tail \citep{2023Natur.616..452L}, similar to those observed in active asteroids whose activation mechanism is considered to be linked to natural impacts \citep{2022arXiv220301397J,2023MNRAS.522.2403T}. Estimates of particle sizes in the ejecta range from micrometers to a few centimeters \citep{2023Natur.616..452L,2023PSJ.....4..138M,2023arXiv230605908R}. Our dynamical Monte Carlo model applied to HST and ground-based images provided constraints on the mass and ejection speed of the particles \citep{2023PSJ.....4..138M}. This model, however, cannot constrain precisely the total amount of mass ejected because it is very insensitive to the boulder population, as those boulders carry a lot of mass, but contribute very little to the brightness. 
In a recent study, \cite{2023ApJ...952L..12J} (hereafter J+23), provided observations of a collection of $\sim$40 meter-sized boulders moving at average projected speeds of $v \sim$0.3 m s$^{-1}$, slightly above the escape velocity of the binary system. Such speeds are about a factor of 50-100 smaller than those observed by LICIACube, constituting clearly a different population. In this paper, we describe this slow-moving boulder population by using dynamical codes, taking advantage of our Monte Carlo models already applied to retrieve the physical properties of the small-sized ejecta particles \citep{2023PSJ.....4..138M}. We analyze the different fates of the ejected boulders, tracking those that may impact on the surfaces of Didymos or Dimorphos, escape the system, or are injected into quasi-stable orbits that might persist during very long times around the binary system, even lasting to the predicted arrival time of the ESA/Hera mission in late 2026.

\section{Dynamical modeling}\label{sec:model}

Our recent Monte Carlo modeling \citep{2023PSJ.....4..138M} provided us with a physical characterization of the ejected particles and the geometry of the launching site compatible with a series of images taken from the ground and the HST \citep{2023Natur.616..452L}. We were able to constrain the particle size distribution function, the ejection velocities, and the geometric parameters of the ejecta cone by performing fits to the observed tail brightness from images acquired from the impact time until late December, 2022. Details of the observational circumstances corresponding to those images are given in Tables 2 and 3 in \cite{2023PSJ.....4..138M}.  However, as indicated in Section \ref{sec:intro}, the boulder distribution cannot be properly modeled as they contribute very little to the observed brightness. Only very deep imaging using sophisticated instrumentation such us the Wide Field Camera 3 (WFC3) on board HST can give the necessary signal-to-noise ratio as to detect confidently the brightness of meter-sized objects at moderate heliocentric and geocentric distances.  Thus, using the aforementioned facility, J+23 were able to detect about forty meter-sized boulders on December 2022 and February 2023 moving around the binary system at projected distances of several thousands kilometers. Owing to the less favourable viewing geometry in February 2023, only the December 2022 image shows clearly all of those boulders, whose positions in the sky are provided by J+23 in their Table 2. That image was a composite of 24 images of 193 s exposure time each obtained through the broadband F350LP filter on 2022 December 19 from 15:05 to 20:25 UT (see J+23, their Table 1).

In \cite{2023PSJ.....4..138M} we obtained ejection speeds, from the simulation of the tails, ranging from about Dimorphos escape speed of $\sim$0.09 m s$^{-1}$ up to 0.375 $r^{-0.5}$ m s$^{-1}$, where $r$ is particle radius in meters  \citep[cf. Table 4 in][]{2023PSJ.....4..138M}. For meter-sized boulders this upper limit would correspond to 0.375 m s$^{-1}$, which is in line with the average projected velocity found by J+23 from the boulders positions in the sky plane. In our simulations, we have selected initially a range of ejection speeds in the 0.09 to 0.7 m s$^{-1}$  interval, in line with the projected speed distribution found by 
J+23. We assume that the ejected boulders came out from the ejecta cone that was seen on the in-situ (cf. LICIACube images) and space-borne and ground-based images. The ejecta cone geometric parameters, i.e., the opening angle and thickness of the cone wall, and the ejecta cone axis direction are given in Table \ref{tab:ejectionpar}. We have updated the values of the ejecta cone direction and cone aperture from \cite{2023Natur.616..457C} and \cite{Hirabayashi2023}, respectively, with respect to the values used in 
\cite{2023PSJ.....4..138M}. In particular, for the ejecta cone opening angle we adopted the mean value of the wide ($\theta_1$) and narrow ($\theta_2$) angle cone parameters obtained by 
\cite{Hirabayashi2023} ($\theta_1$=133$^\circ$ and $\theta_2$=95$^\circ$), i.e. $\theta$=114$^\circ$. The boulder radii are assumed to be logarithmically and randomly distributed between 0.1 and 10 m. Following 
\cite{2023Natur.616..443D}, the best meteoritic analogues for Didymos/Dimorphos are L and LL chondrites, which have densities of about 3500 kg m$^{-3}$, that we assumed an appropriate value for the boulders' density. This value differs from the Design Reference Asteroid (DRA; DART mission internal document) value used for Didymos or Dimorphos (2750 and 2400 kg m$^{-3}$, respectively), implying a bulk density of about 30\% for both bodies \citep{2023Natur.616..443D}. We further assume that the boulders are perfectly absorbing spherical objects, so that the acceleration caused by the radiation pressure force acting on them can be estimated from \citep[e.g.,][]{1968ApJ...154..327F}:
\begin{equation}
    a_{rad}=\frac{Q_{pr} E_s}{c}\frac{3}{8\pi R_h^2 \rho D}
\end{equation}

\noindent
where $Q_{pr}$ is the scattering efficiency for radiation pressure, taken as the unity, $E_s$=3.9$\times$10$^{26}$ W is the total power radiated by the Sun, $c$=3$\times$10$^8$ m s$^{-1}$  is the speed of light, $R_h$ is the heliocentric distance, $\rho$ is the boulder density, and $D$ its diameter.

\subsection{Short-term orbital evolution: Modeling of 
  Jewitt et al. (2023) measurements}

To calculate the boulder orbits we followed two different approaches. On one hand, we used the \texttt{MERCURY} code \citep{1999MNRAS.304..793C}, in which the Sun, Didymos, Dimorphos, and the boulders were all set as massive bodies in the corresponding input file. Although the solar radiation pressure force should be unimportant on such large particles in the short-term, we included such radiation force as it influences the long-term integrations (described below in Section 2.2). This force is included in the user force routine within the code. The integrations are performed by the Bulirsch-Stoer method.  The radii of Didymos and Dimorphos are calculated from the equivalent radii obtained from the ellipsoidal axes taken from the current DRA version 5.20 using the corresponding SPICE kernels (see Table \ref{tab:phys}). The resulting masses of the two components of the binary system are calculated using a common density of 2760 kg m$^{-3}$, in line with the latest DRA solution for Didymos as stated above.  
The orbits of all the bodies are initially propagated from the impact time until 2022 December 19th, the date of the HST image by 
J+23, i.e., 83 days of integration time. The code is initialized by the  Heliocentric Ecliptic (J2000) coordinates and velocities (hereafter HEJ2000 reference frame) of Didymos and Dimorphos taken from the JPL Horizons web interface\footnote{  
https://ssd.jpl.nasa.gov/horizons/}. Each boulder is placed on the impact site location on Dimorphos surface and then a random direction within the ejecta cone and a random speed within the considered limits were set (see Table \ref{tab:ejectionpar}). 

\begin{table}
    \centering
    \caption{Boulder physical properties and ejection parameters.}
    \label{tab:ejectionpar}
    \begin{tabular}{|l|l|}
    \hline
    Radius (m) & 0.1-10, random distribution \\
    Density (kg m$^{-3}$) & 3500 \\
    Ejection speed (m s$^{-1}$) & 0.09-0.7, random distribution  \\
    Cone aperture (deg) & 114\footnote{Mean value of wide and narrow angle ejecta cone from \cite{Hirabayashi2023}}  \\
    Cone wall thickness (deg) & 10 \\
    Ejecta cone direction\footnote{
     Earth Mean Equator J2000 frame \citep{2023Natur.616..457C}} & (-0.72410,0.65198,0.22495) \\
     Impact time (UT) & 2022 September 26, 23:14:24 \\
     Impact site latitude and longitude (deg) & 8.84 S, 264 E\footnote{Dimorphos fixed frame \citep{2023Natur.616..443D}} \\ 
     \hline
     \end{tabular}
  \end{table}

The \texttt{MERCURY} code allows to compute the boulder orbits, but it is only applicable to point masses. Neither Didymos nor Dimorphos are spherical, so that one might wonder what is the effect of assuming non-spherical gravity in the orbital calculations. For that purpose, we started a more sophisticated approach by considering the equation of motion of a dust particle in the Didymos-Dimorphos-Sun system \citep[equation 3 in][]{2023PSJ.....4..138M}, but improving the Didymos and Dimorphos gravity fields terms from spherical to ellipsoidal approximation. To compute those ellipsoidal  acceleration terms we followed the procedure by \cite{Willis2016}, who express the gravity in terms of Carlson symmetric forms of elliptic integrals. For a tri-axial ellipsoidal body, of constant density $\rho$, with  semiaxes $a\ge b\ge c$, the acceleration terms are given by:
\begin{eqnarray}
  a_x=-\mu x R_D(b^2+\kappa_0,c^2+\kappa_0,a^2+\kappa_0)\\
  a_y=-\mu y R_D(a^2+\kappa_0,c^2+\kappa_0,b^2+\kappa_0)\\
  a_z=-\mu z R_D(a^2+\kappa_0,b^2+\kappa_0,c^2+\kappa_0)
\end{eqnarray}

\noindent
where $(x,y,z)$ refer to the coordinates in the corresponding body frame, 
$\mu = G\rho (4/3)\pi a b c$ is the gravitational parameter ($G$=6.6743$\times$10$^{-11}$ kg m$^{-3}$  s$^{-2}$ is the universal gravitational constant), $\kappa_0$ is the largest root of the confocal ellipsoid $C(\kappa)$ defined as:
\begin{equation}
    C(\kappa)=\frac{x^2}{a^2+\kappa} + \frac{y^2}{b^2+\kappa} + \frac{z^2}{c^2+\kappa} - 1,
\end{equation}

\noindent
and $R_D(x,y,z)$ is the Carlson elliptical integral of the second kind.  The equation of motion of the boulders is integrated by a fourth-order Runge-Kutta (\texttt{RK4}) method.  We used NASA’s Navigation and Ancillary Information Facility (NAIF) kernel files and  SPICE toolkit routines to compute the position, velocity, and orientation of Didymos and Dimorphos at each time step. These kernel files are provided by the DART Team, and match the current DRA. In the course of the integration process, we checked whether a collision event with either Didymos or Dimorphos takes place when the following condition holds:

\begin{equation}
    \frac{x^2}{a^2} + \frac{y^2}{b^2} + \frac{z^2}{c^2} \le 1
\end{equation}

\noindent
where $(a,b,c)$ are the semiaxes of the corresponding ellipsoidal body.
As with \texttt{MERCURY}, the integrations are propagated from the impact time until 2022 December 29th. 

\begin{table}
    \centering
    \caption{Assumed physical properties of the binary components.}
    \label{tab:phys}
    \begin{tabular}{|c|c|c|}
    \hline
    Body & Ellipsoid dimensions & Mass  \\
        & ($a\times b\times c$, m) & (kg) \\ \hline
    Didymos &   391.7$\times$391.7$\times$292.0 & 5.18$\times$10$^{11}$ \\    
    Dimorphos & 90.8$\times$69.3$\times$61.3 & 4.46$\times$10$^9$ \\
     \hline
     \end{tabular}
     \end{table}

\begin{figure}
\includegraphics[angle=0,width=0.99\columnwidth]{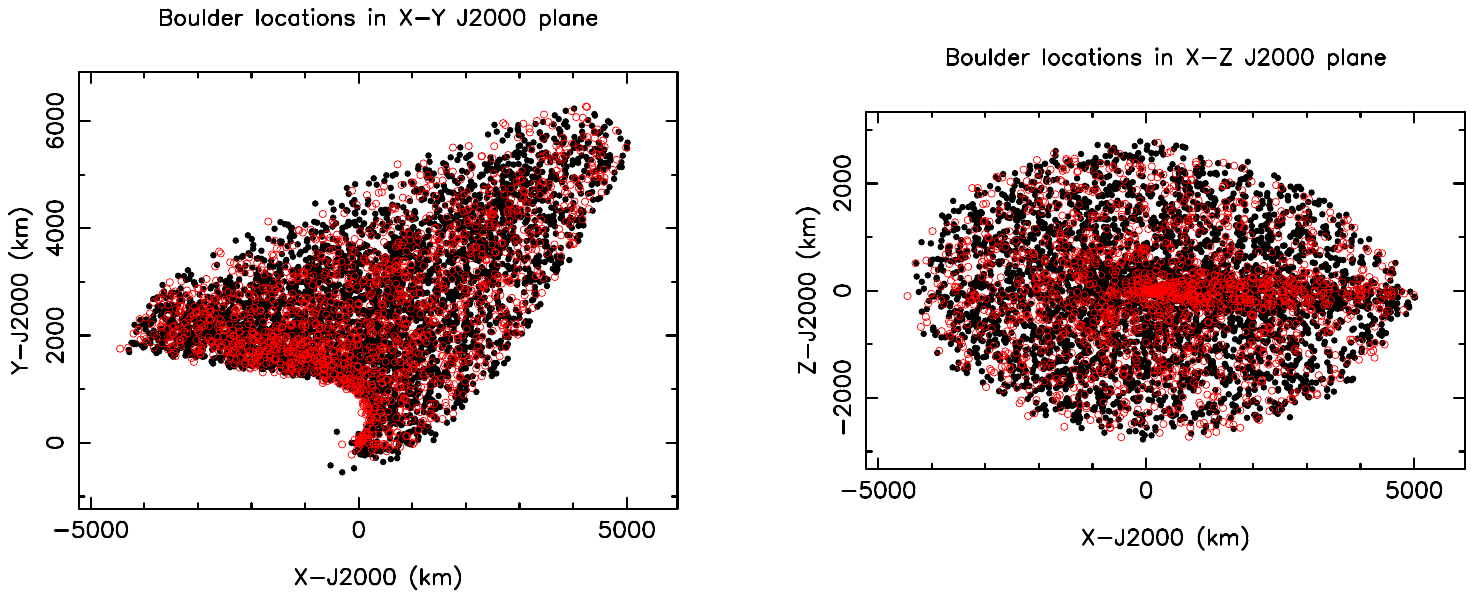}  \caption{The loci of the boulders computed by \texttt{MERCURY} (black dots) and \texttt{RK4} codes (small red open circles) in the HEJ2000 X-Y and X-Z planes on 2022 December 29th. The speed range was 0.09 to 0.7 m s$^{-1}$. 
\label{fig:helioc_comparison}}
\end{figure}

The results of both dynamical codes for the initial random velocity vectors within the imposed limits are compared in Figure \ref{fig:helioc_comparison}, where the locations of the boulders in the HEJ2000 X-Y and X-Z planes are compared. For this comparison, 5000 sample boulders were integrated.

\begin{table}
    \centering
    \caption{Fate of boulders till 2022 December 29th (83 days since impact).}
    \label{tab:fateJewitt}
    \begin{tabular}{|l|l|l|}
    \hline
    & \texttt{MERCURY} &   \texttt{RK4} \\
EVENT TYPE     & code & SPICE \\ \hline
Didymos collision & 7.2\% & 7.2\% \\
Dimorphos collision & 4.0\% & 5.8\% \\
Unbound orbit & 85.9\% & 84.3\% \\ \hline
Bound orbit R$_H\le$ 70 km & 2.9\% & 2.7\% \\
     \hline
     \end{tabular}
     \end{table}
As can be seen, the results by both codes are in very good agreement as they fill the same space in the X-Y and X-Z planes in the HEJ2000 reference frame (see Figure \ref{fig:helioc_comparison}). Only very little differences are found when we analyze the fate of the ejected boulders. The percentage of boulders colliding with Dimorphos or Didymos, those that end up in unbound orbits (i.e., those plotted in Figure \ref{fig:helioc_comparison}), and those that remain in bound orbits (i.e. within the Hill radius of the system, R$_H \le$ 70 km) are given in Table \ref{tab:fateJewitt}. The highest divergence between models occurs for those boulders colliding with Dimorphos, whose probability turns out to be a bit higher (5.8\%) for \texttt{RK4} than for \texttt{MERCURY} (4.0\%). 

Despite Didymos having a larger cross section than Dimorphos ($\times 25$), Didymos experiences a collision rate just $\sim 1.5$ times higher than Dimorphos (i.e. $\times 17$ less than expected). This can be explained by the fact that particles are ejected from Dimorphos at low relative speeds, causing them to enter the binary system with velocities close to Dimorphos' orbital velocities, after escaping its Hill sphere. As a result, these particles form a torus around Didymos, with periastrons located near Dimorphos' orbit. Consequently, the particles frequently intersect Dimorphos' orbit, increasing the likelihood of collisions. On the other hand, particles colliding with Didymos must either be ejected at higher velocities, or undergo orbital evolution to decrease their periastrons to the size of Didymos' radius.

\begin{figure}
\includegraphics[angle=0,width=0.99\columnwidth]{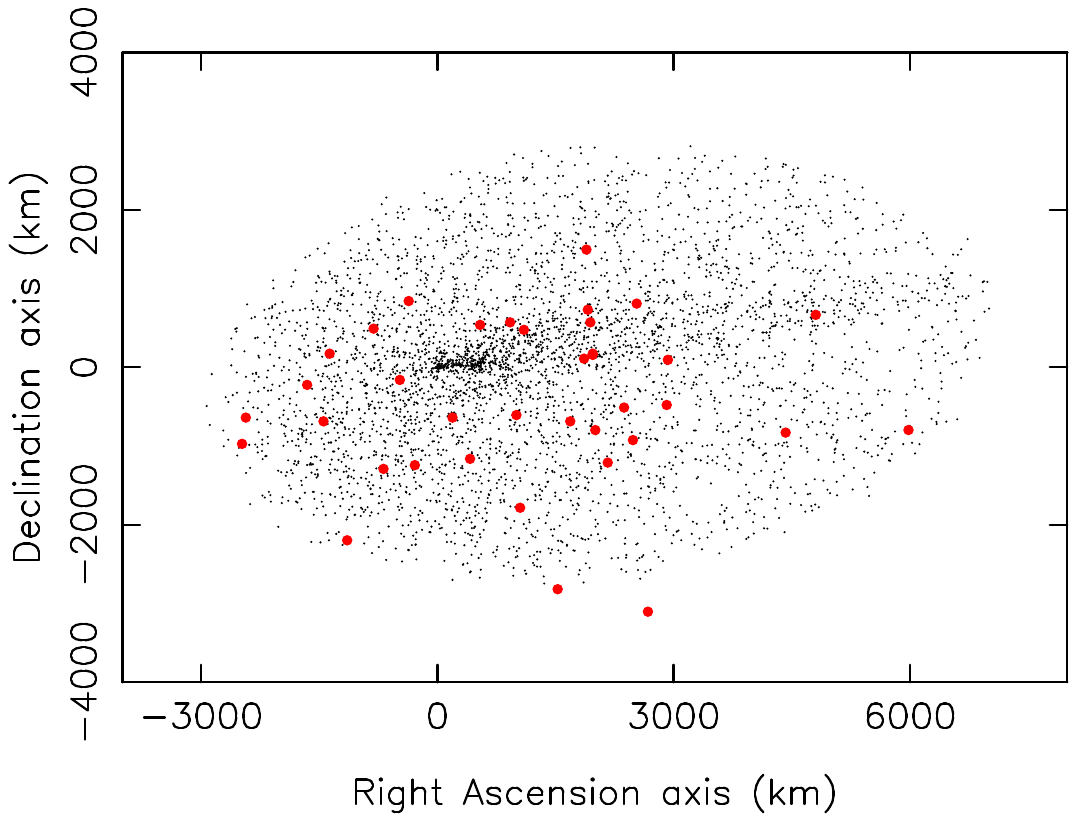}   
\caption{The loci of the boulders observed by 
\cite{2023ApJ...952L..12J} 
on the plane of the sky on 2022 December 29th (red symbols) together with the model calculations of the boulder positions using the \texttt{RK4} code with SPICE kernels for the current 
Design Asteroid Reference (DRA) version 5.20, with the input model parameters of Table \ref{tab:ejectionpar} (black dots).  North is up and East to the left. 
\label{fig:jewitt}}
\end{figure}

The positions of the boulders projected on the sky plane can be obtained from their HEJ2000 coordinates by standard methods. Figure \ref{fig:jewitt} displays the projected positions in the sky plane from their HEJ2000 coordinates (red symbols in Figure \ref{fig:helioc_comparison}), together with the boulder positions found by 
J+23, computed from their Table 2.  As Figure \ref{fig:jewitt} shows, the region of the sky plane covered by the modeled boulders covers reasonably well that of the observed boulders. There is just one boulder out of that region, which was probably ejected at a speed greater than 0.7 m s$^{-1}$, the one located in the southernmost portion of the plot. Note that the boulders observed by  J+23 are all located several hundred kilometers away from the Hill sphere of the binary system. As a result, these boulders have already escaped the system and entered into unbound heliocentric orbits.

\begin{figure}
\includegraphics[angle=0,width=0.99\columnwidth]{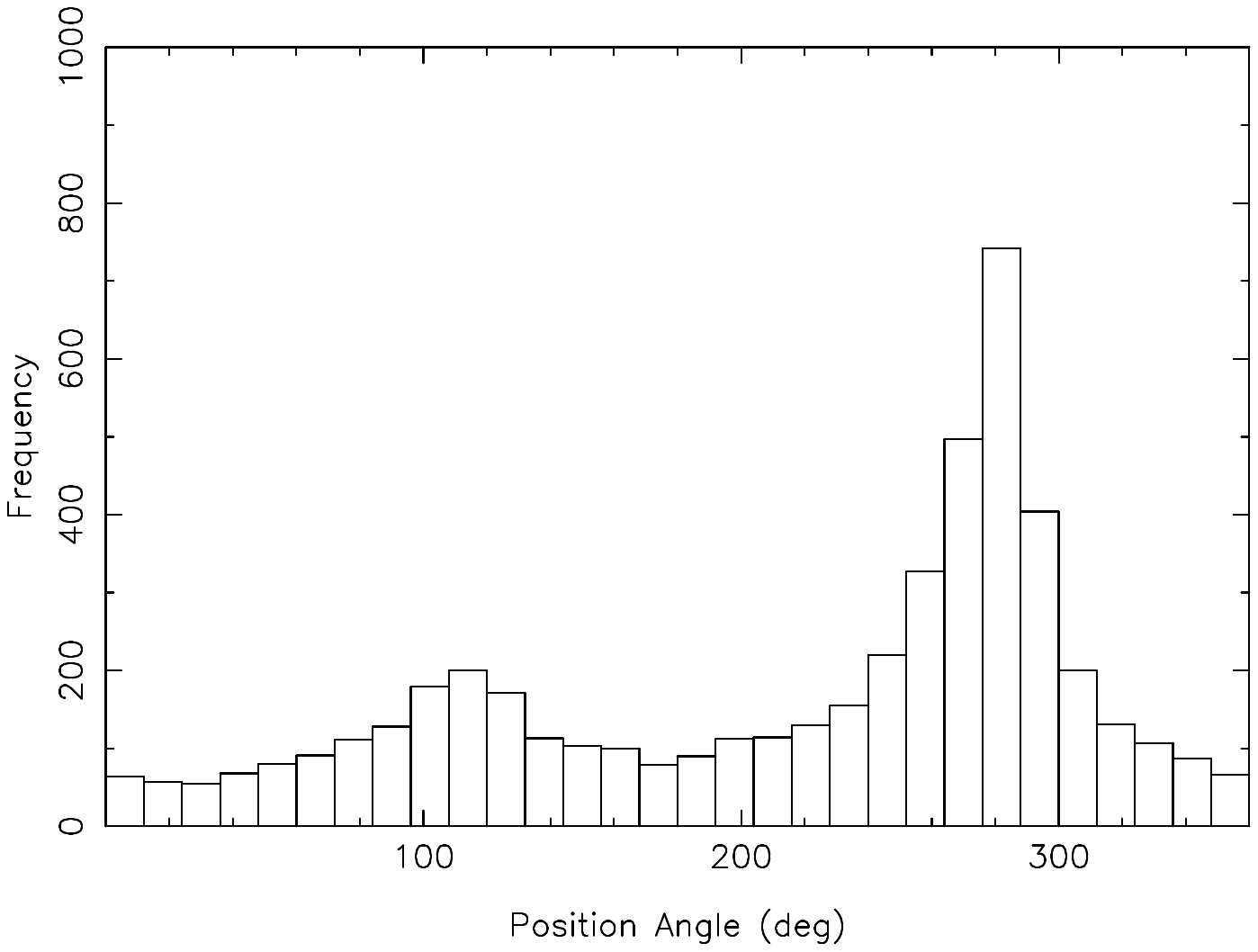}   
\caption{Histogram of the position angles of the modeled boulders (see Figure \ref{fig:jewitt}, upper panel), showing maxima at $\sim$280$^\circ$  and $\sim$120$^\circ$ of position angle. 
\label{fig:PA}}
\end{figure}

As noted by 
J+23, we also find a north-south asymmetry in the modeled boulder locations, the southern part showing a higher boulder surface density than the northern one. In a similar way to that done by J+23 we have also computed a histogram of the distribution of boulders as a function of the position angle (see Figure \ref{fig:PA}) which shows a maximum near 280$^\circ$ and a secondary maximum near 120$^\circ$, in agreement with the results of Figure 4 by J+23.

\begin{figure}
\includegraphics[angle=0,width=0.99\columnwidth]{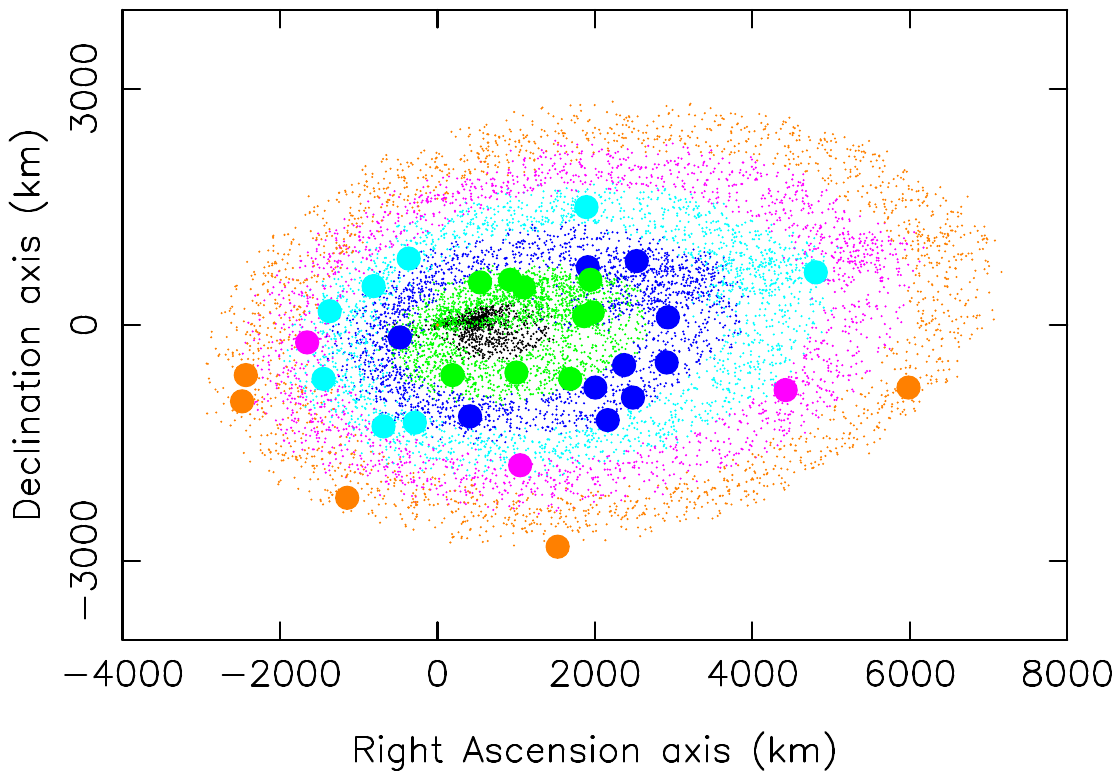}   
\caption{Small dots: The loci of the modeled boulders on the sky plane on 2022 December 29th (see Figure \ref{fig:jewitt}), but sorted in different colors according to the initial ejection speed as described in Table \ref{tab:colorbins}. The large solid circles of different colors correspond to the measured boulder positions by  J+23. 
\label{fig:vel_components_jewitt}}
\end{figure}

The effect of solar radiation pressure on the meter-sized boulders for the integration time of 83 days is small, so the location of the modeled boulders on the sky plane depends mostly on the ejection velocities. Figure \ref{fig:vel_components_jewitt} shows the location of the modeled boulders  in a way similar to Figure \ref{fig:jewitt}, but with different colors according to specific ejection speed bins as shown in Table \ref{tab:colorbins}. The regions on the sky plane occupied by different ejection speed ranges are different, so we can identify each observed boulder by J+23 as belonging to a specific speed bin and assign to it the corresponding color. Table \ref{tab:colorbins} shows the amount of observed boulders in each bin. Since in some cases the observed boulders fall near a border between two speed regions, we just provide a range of boulders instead of a single number. Although the splitting into several speed ranges is very coarse, we can see that most of the boulders appear highly concentrated in the first two bins of the histogram, with ejection speeds smaller than 0.395 m s$^{-1}$. Boulders ejected at speeds smaller than the lower limit of the first bin (0.192 m s$^{-1}$) can not be seen in J+23 figure. Any boulder ejected at those speeds would be hidden in the highly saturated portion of the image near the binary asteroid location. Since a dramatic cutoff in the speed distribution of boulders between escape speed and 0.192 m s$^{-1}$ is not expected, and given the tendency shown in Table \ref{tab:colorbins}, we believe that such range of small speeds will be significantly populated. 

\begin{table}
    \centering
    \caption{Modeled frequency distribution of speeds of the boulders  observed by J+23 and corresponding color code.}
    \label{tab:colorbins}
    \begin{tabular}{|l|l|l|}
    \hline
Speed bin  & Number of &  Color code \\
(m s$^{-1}$) & boulders & \\ \hline
0.090--0.192 & -- & Black \\
0.192--0.293 & 9 & Green \\
0.293--0.395 & 9--10 & Dark blue \\
0.395--0.497 & 7--8 & Light blue \\
0.497--0.598 & 3-4 & Magenta \\
0.598--0.700 & 5  & Brown \\
     \hline
     \end{tabular}
     \end{table}

\begin{figure}
\includegraphics[angle=0,width=0.99\columnwidth]{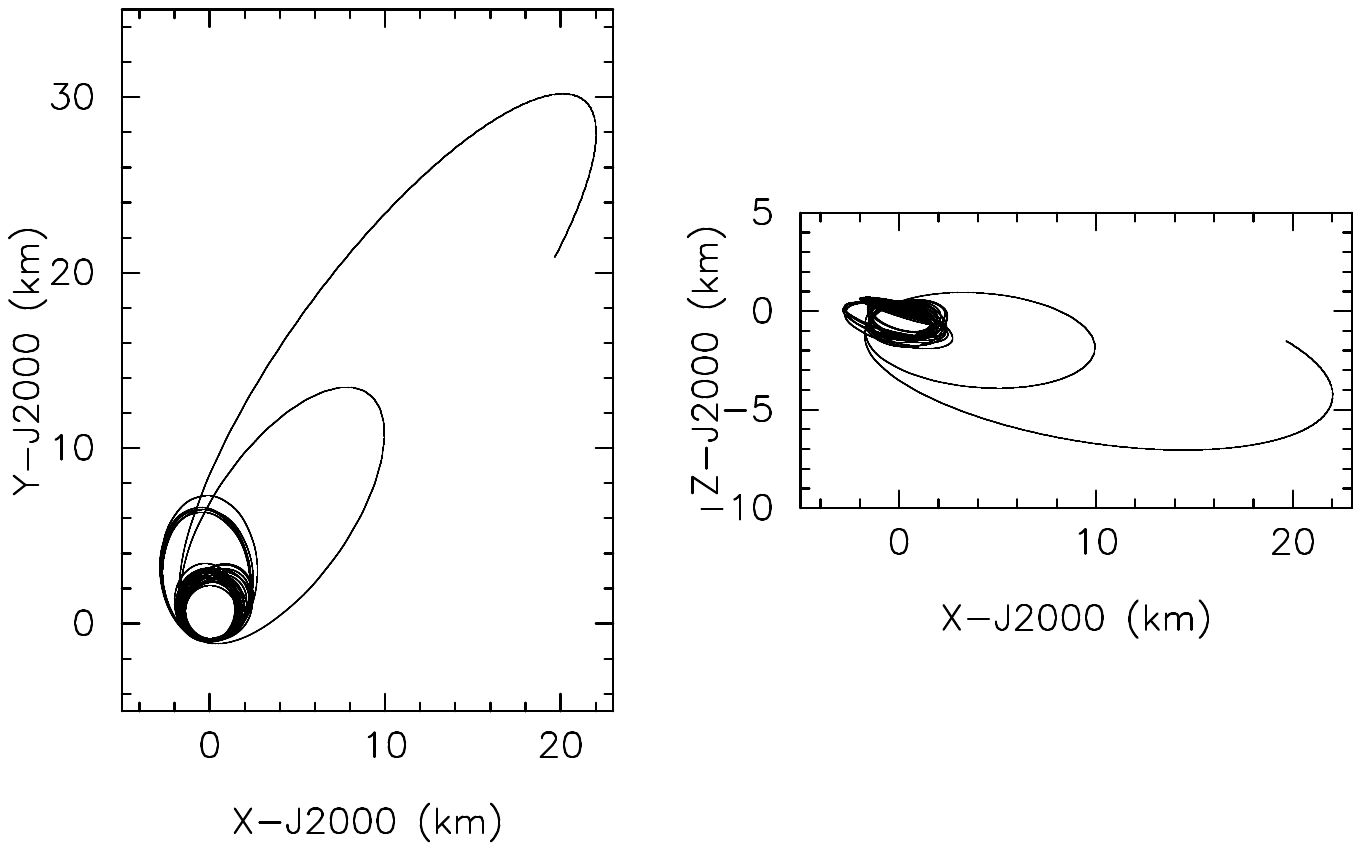}   
\caption{Example of an orbit of a boulder of radius of 7.2 m ejected at a speed of 0.128 m s$^{-1}$ integrated over a period of 83 days since impact. Left panel represents the orbit projected to the X-Y HEJ2000 plane, and the right panel, to the X-Z plane.
\label{fig:exampleorbit83}}
\end{figure}

From the distribution of modeled boulders, we see that a small, but non-negligible, amount (around 3\%, see Table \ref{tab:fateJewitt}) remain in close orbits within the Hill radius of the system after the 83 day integration period. An example of those bound boulders is given in Figure \ref{fig:exampleorbit83}. All the surviving boulders were found to be ejected at speeds smaller than 0.22 m s$^{-1}$, with a median speed of 0.13 m s$^{-1}$. This upper limit of 0.22 m s$^{-1}$ is in agreement with the theoretical estimate of $v\approx$ 0.24 m s$^{-1}$ for the escape velocity from the binary system at the distance of Dimorphos from Didymos. It is interesting to check whether some of those boulders would survive for longer timescales, as discussed in the next subsections.


\subsection{Long-term orbital evolution of boulders: Assessing the presence of large size debris at the time of the Hera (ESA) mission arrival}
\begin{figure}
\includegraphics[angle=0,width=0.99\columnwidth]{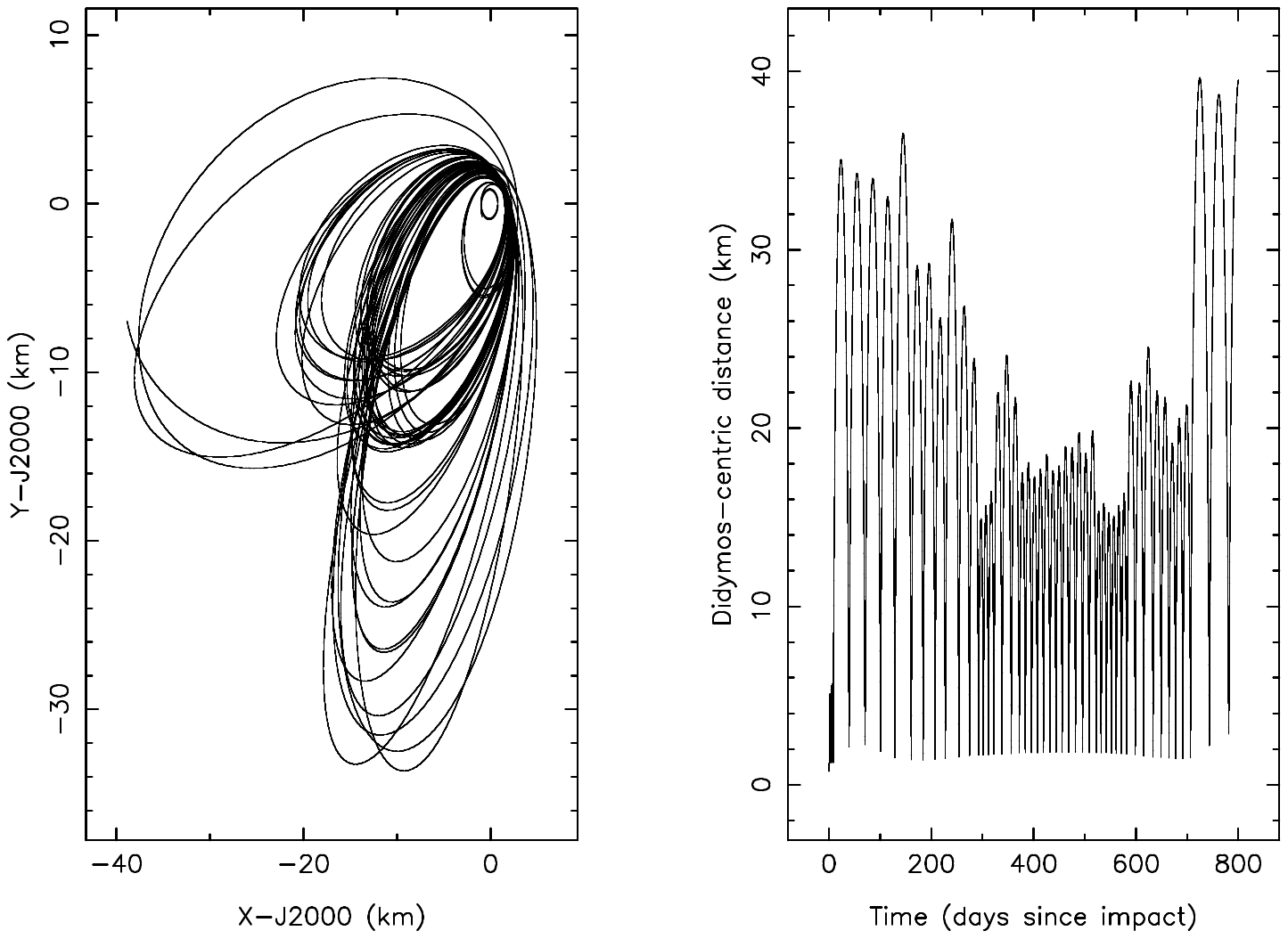}   
\caption{Example of a boulder in close orbit after 800 days of integration time. The boulder radius is 3.8 m, and the ejection velocity is the Dimorphos escape velocity, 0.09 m s$^{-1}$. 
\label{fig:orbit_sample800}}
\end{figure}

The amount of boulders that remain close to the binary system after the 83-day period raises the question of as to how long those boulders could survive orbiting the system. We then integrated the equation of motion for longer time spans, using both dynamical model approaches, up to 800 days (i.e., 2024 November 30). The limitation of 800 days in the orbit propagation is set by the time coverage of Dimorphos ephemerides, which ends on 2024 December 31, using the current SPICE meta-kernel d520\_v02. The initial conditions are set again to the values indicated in Table \ref{tab:ejectionpar}, and  5000 boulders were integrated. Table \ref{tab:fate800} gives the results of those longer integrations, revealing again a good agreement between both modeling techniques.  There is a significant reduction in the bounded boulder population to the 0.3-0.5\% level with respect to the 83-day runs, which evolves with a simultaneous increase in the amount of collisions with Didymos and Dimorphos, as we will see later in this section (see Figure \ref{fig:impact_histograms}). An example of an orbit of a surviving boulder after the 800 days integration time is given in Figure \ref{fig:orbit_sample800}. In general, except at the very beginning of the orbital evolution, the boulder pericenter is always larger than about 2 km, rapidly reaching highly eccentric elliptical orbits with apocenters comparable to the Hill radius.

\begin{figure}
\includegraphics[angle=0,width=0.99\columnwidth]{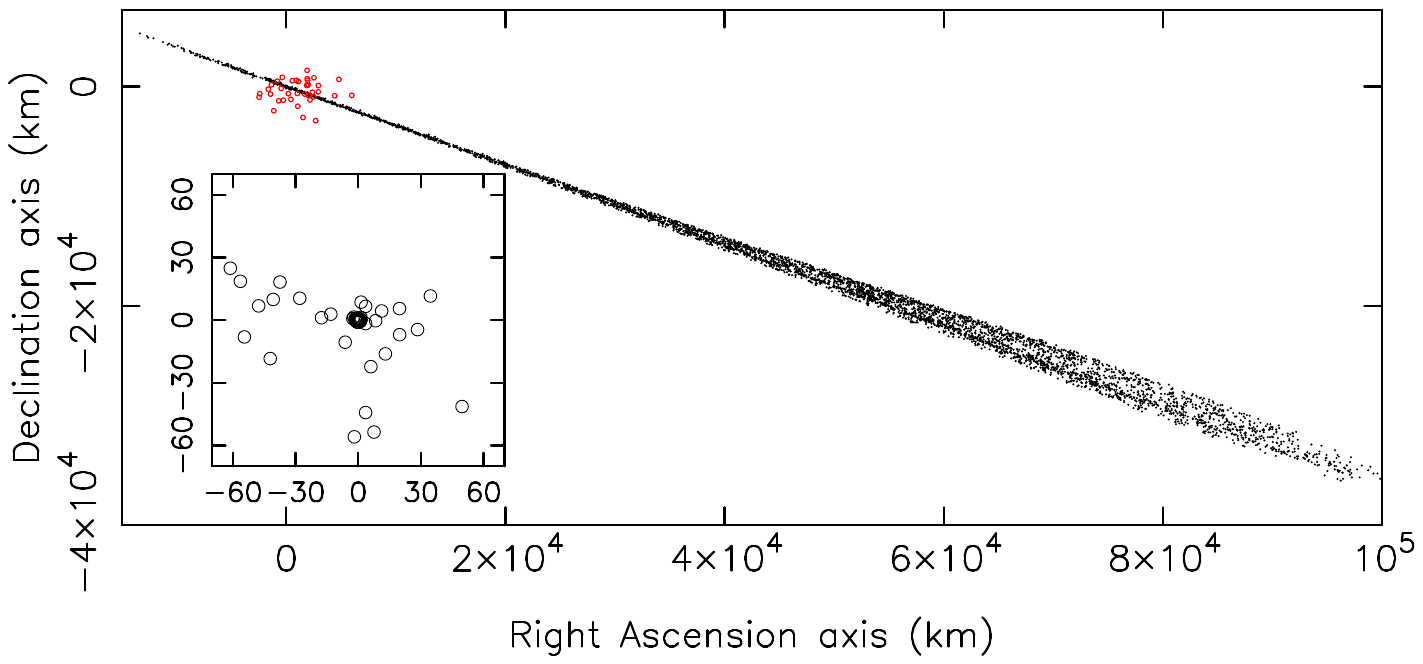}   
\caption{The black dots represent the positions of the modeled boulders on the sky plane after 800 days of orbital evolution, for the ejection parameters of Table \ref{tab:ejectionpar}. The small red open circles are the positions of the boulders observed by \cite{2023ApJ...952L..12J} on 2022 December 19, i.e., 83 days after impact. The inset displays a zoomed region 140$\times$140 km$^2$ around Didymos position, showing the location of the modeled boulders (black open circles) within or near Hill radius. North is up, East to the left in both panels.
\label{fig:loci800}}
\end{figure}

In Figure \ref{fig:loci800}, we plot the position on the sky plane of the boulders that were integrated for 800 days, using the same procedure described above. This region of the sky displays the spatial dispersion of the modeled boulders reached by the combined effect of different ejection velocities and radiation pressure accelerations, and would represents the boulder layout at the beginning of the Didymos observing window starting in late 2024. Also plotted are the positions of the boulders observed by J+23 after 83 days since ejection time for spatial scale comparison purposes. The figure inset is a zoomed region near the asteroid position showing the location of the modeled boulders that remain close or within the Hill radius after the 800-day integration time.

\begin{table}
    \centering
    \caption{Fate of boulders integrated till 800 days after impact.}
    \label{tab:fate800}
    \begin{tabular}{|l|l|l|}
    \hline
     & \texttt{MERCURY} &   \texttt{RK4} \\
EVENT TYPE & code & SPICE \\ \hline
Didymos collision & 8.4\% & 7.9\% \\
Dimorphos collision & 6.1\% & 5.5\% \\
Unbound orbit & 85.2\% & 86.1\% \\ \hline
Bound orbit R$_H\le$ 70 km & 0.3\% & 0.5\% \\
     \hline
     \end{tabular}
     \end{table}

\begin{figure}
\includegraphics[angle=0,width=0.99\columnwidth]{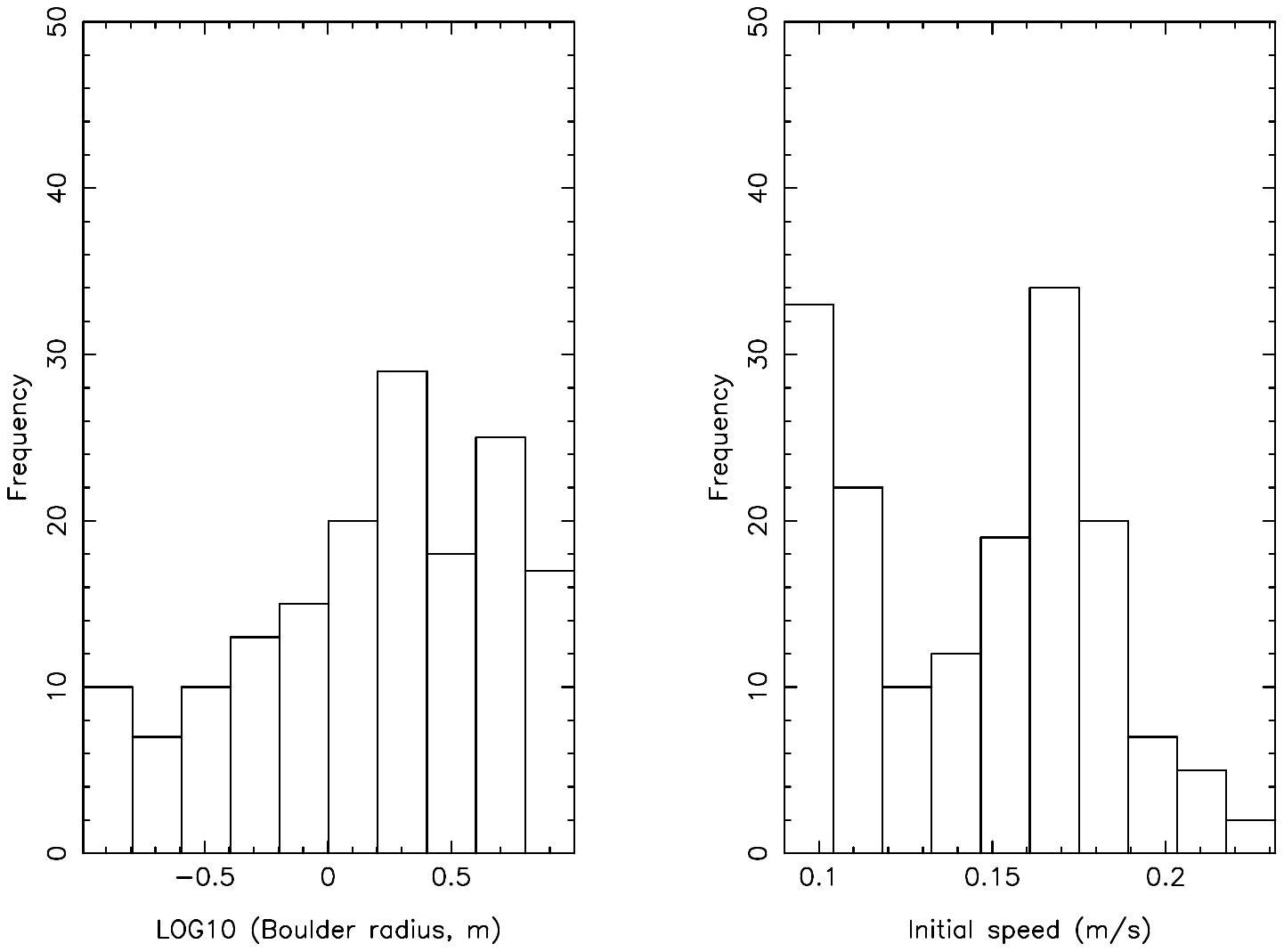}   
\caption{Left panel: Distribution of the radius of the boulders that remain in orbit after 1550 days of integration. Right panel: Distribution of initial velocities of those surviving boulders. 
\label{fig:radii_and_vel_boulders}}
\end{figure}

\begin{figure}
\includegraphics[angle=0,width=0.99\columnwidth]{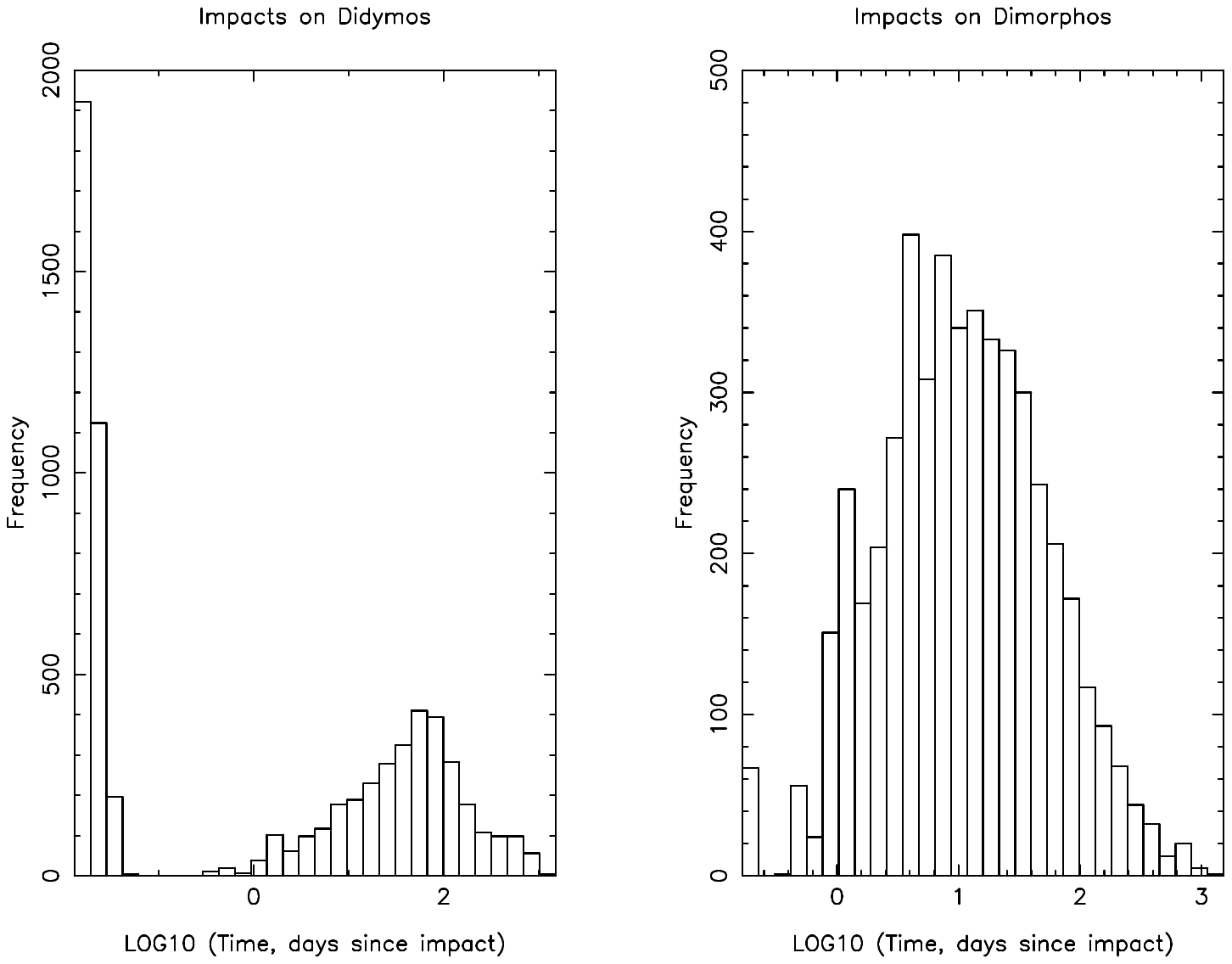}   
\caption{Frequency of boulders impacting on Didymos and Dimorphos versus time. 
\label{fig:impact_histograms}}
\end{figure}

For still longer integration times we rely on the results by the \texttt{MERCURY} code only. Since we have already shown that the consideration of non-spherical gravity does not affect significantly the results obtained, we do not expect detectable differences when using the more precise calculations given by ellipsoidal gravity. Our aim is to perform integration of orbits till the predicted arrival time of the ESA/Hera mission in late 2026, to estimate the percentage of the ejected boulders that after $\sim$1550 days of integration since impact might still remain in orbits close to the binary system.  We also seek to assess the impact rates on the two asteroids during such long time baseline, and to estimate the escape speed of the boulders versus time. Those hypothetical boulders could be observed with on-board instruments by Hera. Owing to the very limited amount of boulders that remained in orbit after the 800-day integration period, and in order to obtain a meaningful statistics for the longer evolution time of $\sim$1550 days, we integrated a much larger amount of boulders than in the previous runs, specifically 76,300, using, as before, the initial conditions shown in Table \ref{tab:ejectionpar}. The statistics concerning the fate of the boulders is given in Table \ref{tab:fate1550} (4-body column). We find a kind of asymptotic limit in the orbital behavior, with very little differences with respect to the 800-day orbital evolution shown in Table \ref{tab:fate800}.  The boulder population in bounded orbits has decreased to the 0.2\% level. Those surviving boulders (only 165 out of the 76,300 ejected) have radii mostly larger than about 1 m, and were ejected with speeds smaller than $\sim$0.22 m s$^{-1}$, as before (see Figure \ref{fig:radii_and_vel_boulders}). Previous studies  
\citep[e.g.][]{2022PSJ.....3..118R} have also confirmed that some ejecta could survive in the binary environment for timescales comparable to the Hera arrival time.

Making a very conservative estimate, let us assume that the 37 boulders spotted by J+23 are the whole 85\% of boulders with unbound orbits at the end of the 1550-day evolution time (see Table \ref{tab:fate1550}): that would make a total of 44 boulders ejected from Dimorphos, according to our model statistics. 
Considering that the 8.4\% of the ejected population  would impact back to Didymos (see Table \ref{tab:fate1550}), the estimated number of such boulders is 3.7. Assuming a Poisson probability distribution, the probability that at least one event takes place when $\lambda$=3.7 events are expected is given by $p$=$\int_{1}^{\infty}P(x) dx$, where $P(x)=\frac{\lambda^x}{x!}e^{-\lambda}$, which gives a probability of $p$=97.5\% that Didymos undergoes at least one impact.  In the case of Dimorphos, the corresponding probability would be $p$=93.2\%. Moreover, we expect a 0.2\% of the total population of boulders being bound to the Didymos system after 1550 days of orbital evolution, i.e., 0.088 boulders, which corresponds to a non-negligible probability of $p$=8.4\% of having at least a boulder orbiting within 70 km of the binary system at the time of the Hera arrival.

\begin{table}
    \centering
\caption{Fate of boulders integrated till 1550 days after impact. All results refer to the \texttt{MERCURY} code.}
    \label{tab:fate1550}
    \begin{tabular}{|l|l|l|l|l|}
    \hline 
EVENT TYPE &  4-body \footnote{Orbit propagation using 4 bodies (see text)} &  9-body \footnote{Orbit propagation using 9 bodies} & $v$=0.09--0.2 m s$^{-1}$ \footnote{Orbit propagation with 9 bodies, but with restricted speed range, as indicated}& 
     $r$=0.1 m \footnote{Orbit propagation with 9 bodies, but limiting the boulder size to $r$=0.1 m} \\ \hline
Didymos collision & 8.6\% & 8.4\%  & 19.1\% &  10.6\%      \\
Dimorphos collision & 6.5\% & 6.1\%  & 31.5\% & 6.7\%        \\
Unbound orbit & 84.7\%  & 85.3\% & 48.7\%   &  82.6\%      \\ \hline
Bound orbit R$_H\le$ 70 km & 0.2\% & 0.2\% & 0.7\% & 0.1\%  \\
     \hline
     \end{tabular}
     \end{table}

\begin{figure}
\includegraphics[angle=0,width=0.99\columnwidth]{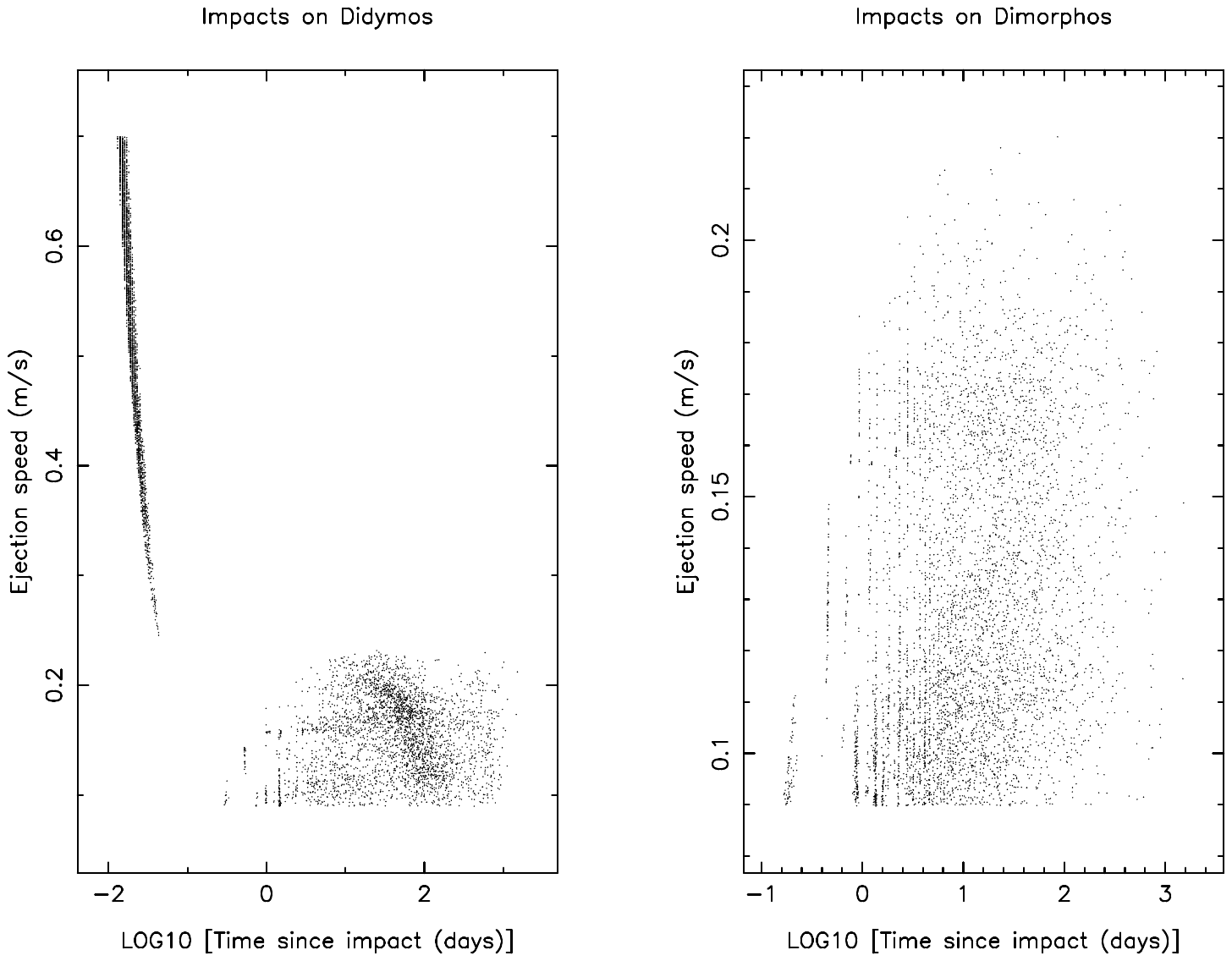}   \caption{Ejection speeds as a function of the impact time of the modeled boulders on Didymos and Dimorphos. 
\label{fig:impact_init_speeds}}
\end{figure}

A histogram of impact frequency on Didymos and Dimorphos is given in Figure \ref{fig:impact_histograms}. Impacting times on Didymos and Dimorphos as a function of the ejection speed are given in Figure \ref{fig:impact_init_speeds}. The frequency distributions of the impacting boulders is markedly different on Didymos than on Dimorphos. On Didymos, after a pronounced maximum of impact rate within the first hour (the maximum is around 24 minutes after DART impact), the frequency suddenly decreases and then smoothly starts to increase after $\sim$10-20 days, up to a secondary maximum centered $\sim$100 days after impact, then decreases smoothly again. Concerning the first prominent maximum, most of those impacting boulders have relatively large ejection speeds (see Figure \ref{fig:impact_init_speeds}), greater than $\sim$0.5 m s$^{-1}$ in most cases. From the measurements by 
J+23, we see that the abundance of such relatively high-speed boulders is presumably small (see Table \ref{tab:colorbins}), so that the actual frequency of early impacting boulders on Didymos might be greatly reduced. In fact, photometric measurements during those first minutes after DART impact do not reveal any significant brightness enhancement \citep[see e.g.][]{2023PSJWeaver,2023Natur.616..461G} that could be attributed to re-impacting boulders. On the contrary, at later times, all the remaining impact events on Didymos and Dimorphos are associated to ejection speeds smaller that $\sim$0.22 m s$^{-1}$ (see Figure \ref{fig:impact_init_speeds}), and are therefore much more likely events from the measurements by J+23.

Regarding Dimorphos, after a small, short-lasting avalanche, the impact rate starts to increase, reaching a broad maximum centered at $\sim$10 days, followed by a decrease. This is consistent with our statement that a possibility of the brightness increase after 7-8 days after impact \citep[the so-called 8-day photometric bump, see e.g.][]{Kareta2023, 2023arXiv231101982R, 2023arXiv231101971M} might be explained by re-impacting debris \citep[see][]{2023PSJ.....4..138M}. Also, the generation of the secondary tail, which appeared at about the same dates \citep[$\sim$6 days after impact, see][]{2023Natur.616..452L} might be related to the same phenomenon, although there are also other mechanisms that could be playing a role in this context \citep[see][]{Ferrari2023, 2023ApJ...956L..26K}.

\begin{figure}
\includegraphics[angle=0,width=0.99\columnwidth]{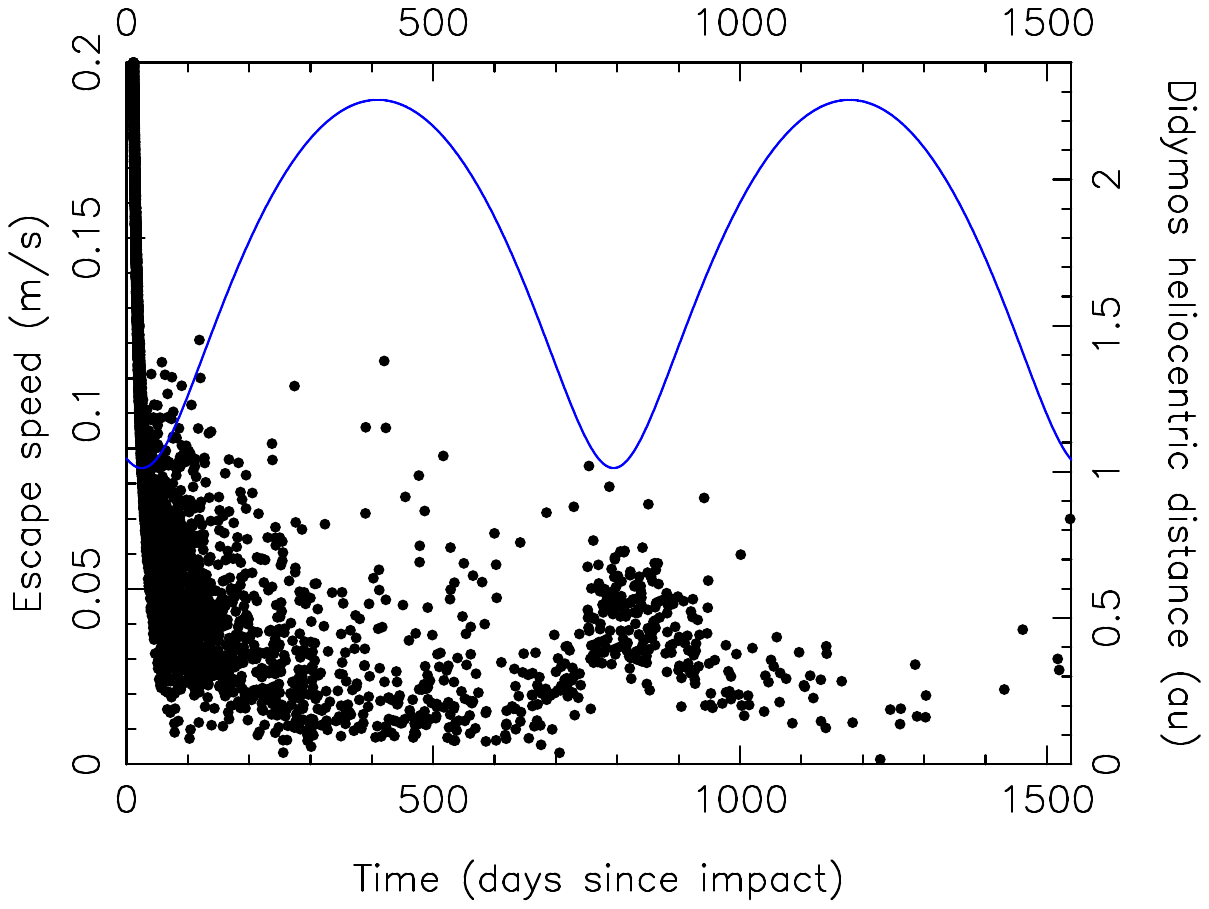}   
\caption{Escape speeds (speeds calculated at 200 km from Didymos) of the boulders ejected (solid circles, left axis). The blue line (right axis) shows the heliocentric distance of Didymos. A clear anticorrelation between escape speeds and heliocentric distance is noticed. 
\label{fig:escape_speed}}
\end{figure}

We have also computed the distribution of the boulder escape speeds, calculated at radius vectors of $r$=200 km, as a function of time. This distribution is shown in Figure \ref{fig:escape_speed}, together with Didymos heliocentric distance. We see a clear anticorrelation between the escape speeds and the heliocentric distance. Escape speeds are higher for shorter heliocentric distances, as expected. Additionally, we have observed a correlation between the number of escapes and Didymos' heliocentric distances. Following the high rate of escapes in the days immediately after the impact, which coincided with the perihelion passage, the rate decreases during the aphelion and then increases once again during the subsequent perihelion.

As a further check of the model, we have also calculated the orbital evolution of the boulders using \texttt{MERCURY} for the long integration time of 1550 days, and with the same input settings as before (see Table \ref{tab:ejectionpar}), but including all the inner planets and Jupiter as perturbing bodies. Owing to the longer CPU time respect to the 4-body integration, we reduced the number of boulders integrated to 40,000, but we have verified that it does not make any difference as far as the resulting statistics, by comparing the results of the 4-body problem using the same amount of 40,000 boulders. The different boulder fates percentages are shown in Table \ref{tab:fate1550}, column labeled as 9-body, where the results are essentially coincident with those corresponding to the 4-body problem,  demonstrating that the gravitational influences of the inner planets plus Jupiter on the dynamics of the ejected boulders is negligible, at least for the mentioned time interval.

Concerning the input model parameters, we have performed  sensitivity tests for two parameters that might influence the outcome of the models: the boulder speeds, and their size, in the former case because it determines the rate of impacts with the two binary components, and, in the latter because the radiation pressure might play a role in the long term evolution, although a much smaller effect is expected. 
The most influential parameter by far is the boulder ejection speed. In Section 2.1, we indicated the possibility, based on J+23 data, that the  population of boulders being ejected at speeds larger that Dimorphos escape velocity but smaller than $\sim$0.2 m s$^{-1}$ might be significant. To study the sensitivity of the model to the ejection speeds, we rerun the previous long-term evolution with all 9 bodies involved, but now limiting the ejection speeds to the range 0.09 to 0.2 m s$^{-1}$. The results are displayed in Table \ref{tab:fate1550}, fourth column. The percentages of probability of collision with Didymos and Dimorphos have increased notably with respect to the nominal model (third column). Moreover, the probability of collision with Dimorphos is now significantly higher than that with Didymos, owing to the smaller speeds assumed. Regarding the boulder population at the end of the integration, it has increased to 0.7\%, i.e., more than three times higher than that of the nominal model. With respect to the run using only the smallest size of the distribution, $r$=0.1 m, the results (see last column of Table \ref{tab:fate1550}) are not very different to those using the full size distribution, except for the amount of surviving boulders in bound orbits, that has decreased by 50\%, by the effect of radiation pressure, as expected.

\section{Discussion and Conclusions}

In situ observations with instrumentation on-board LICIACube, and remote, using the HST WFC3, have shown the presence of significant populations of meter-sized boulders that have been ejected after the impact of DART on Dimorphos' surface. While the boulders observed by LICIACube move at speeds of several 10 m s$^{-1}$, those imaged by WFC3 had projected speeds of $\sim$0.3 m s$^{-1}$. Starting from the physical parameters of the small-sized ejecta, we analyzed this slow-moving boulder population by using dynamical codes. We propagated the orbits of a statistical sample of boulders until 83 days, i.e., the observation time in the image by J+23
and, assuming the model parameters of Table \ref{tab:ejectionpar}, we have estimated the statistical distribution of the speeds. We have also found a strong similarity between the position angle distribution of the observed boulders in the sky plane and that found from the dynamical model. From the modeled distribution of ejected boulders we derive that about 3\% of all the ejected boulders remain in orbit at the end of the 83-day integration time. In this regard, \cite{2018Icar..312..128Y} performed pre-impact simulations of the evolution of the ejecta using a similar simulation time of 60 days. Leaving apart the differences in some of the input parameters, such as the ejection pattern, they found a 5\% of 10 cm sized particles ejected between 0.12 and 0.18 m s$^{-1}$ that remain orbiting the system after that 60-day period, which is in line with what we found after the 83-day of orbital evolution. 

To check the fate of the boulders on longer timescales, we performed  integrations lasting 800 and 1550 days after impact, which, in the latter case, corresponds to the approximate time of the Hera arrival in late 2026. Those runs show that the amount of bounded boulders are reduced with time, mostly by collision with Didymos and Dimorphos, although the relative percentages of the different fates do not vary much after 800 days. 
The amount of bounded boulders after 1550 days is just $\sim$0.2\%, all of them being ejected at speeds smaller than $\sim$0.22 m s$^{-1}$. Since this population of very slow boulders is not constrained by the observations made by J+23, as they would be hidden by the glare in the saturated portion near the asteroids optocenter by CCD blooming on the WFC3 HST images, we cannot give a firm estimate of the absolute number of boulders that might be close to the binary system when Hera arrives. However, assuming that the boulder population spotted by J+23 corresponds to the total number of unbound boulders, we provide a very conservative estimate of a 8.4\% probability of having at least a boulder bound to the binary system at the time of the Hera arrival. Regarding the long-term evolution of boulders, \cite{2022PSJ.....3..118R} also estimated that a certain number of large particles, in the 5 to 10 cm size range, ejected with speeds around 0.1 m s$^{-1}$, might survive for several years, bringing up the possibility that some of them could remain in orbit when Hera arrives.

Concerning the probability of collisions of boulders with the binary components, the percentages turn out to be $\sim$8\% and $\sim$6\% on Didymos and Dimorphos, respectively, for the input parameters of Table \ref{tab:ejectionpar}. Regarding Didymos, 
we observed a pronounced maximum of impact rate within the first day after DART impact, and after a short lapse, a smooth increase is seen starting after $\sim$10-20 days, up to a secondary maximum centered $\sim$100 days after impact. On Dimorphos, after a sharp maximum right after the impact, a broader maximum develops peaking $\sim$10 days after impact. 
The occurrence of impacts by boulders on Didymos is of particular interest for Hera. In fact, Didymos is spinning close to its stability limit, and any perturbation on its surface may trigger landslides and modification of surface features. The landscape shown by the Didymos Reconnaissance and Asteroid Camera for Optical navigation (DRACO) images may therefore differ from what the Hera spacecraft will show. That circumstance may hinder the reconstruction of the Didymos surface age determination. On the other hand, and taking into account the large boulder masses, the impinging momenta might be significant, triggering perhaps the ejection of small size debris from both bodies, that might be feeding the tail episodically, inducing short-lived brightening events, such as that detected 7-8 days after impact (the so-called 7-8 days photometric bump). Unfortunately, given the fact that the angle between the Earth and Didymos orbital plane is quite small during the coming months (and at least until the end of 2026), the emergence of secondary tails produced by such small-sized debris, along synchrones different to that corresponding to the impact time, will be very difficult to observe, as they will be all overlapped. Finally, we raise as a possibility that this evenly distribution of impacting boulders phenomenon can also lead to the emergence of transient dust tails in natural active asteroids, increasing a bit more the complexity of activation mechanisms from such remarkable objects.


\newpage
\section{Acknowledgments}

 This work was supported by the DART mission, NASA Contract 80MSFC20D0004.
 
FM acknowledges financial supports from grants PID2021-123370OB-I00, and from the Severo Ochoa grant CEX2021-001131-S funded by MCIN/AEI/ 10.13039/501100011033.

GT acknowledges financial support from project FCE-1-2019-1-156451 of the Agencia Nacional de Investigaci\'on e Innovaci\'on ANII and Grupos I+D 2022 CSIC-Udelar (Uruguay).

ACB acknowledges funding by Spanish Government MICINN project (PGC 2021) PID2021-125883NB-C21.


%

\vspace{5mm}

\bibliography{sample631}{}
\bibliographystyle{aasjournal}



\end{document}